# Optimally shaped narrowband pulses for Femtosecond Stimulated Raman Spectroscopy in the range 330-750 nm

E. Pontecorvo[1], C. Ferrante[1], C. G. Elles[2] and T. Scopigno[1*]

[1] *Dipartimento di Fisica, Universita' Roma "Sapienza", P.le Aldo Moro 2, Roma, Italy*
[2] *Department of Chemistry, University of Kansas, Lawrence, Kansas 66045, United States*
[*] *tullio.scopigno@phys.uniroma1.it*

**Abstract:** Spectral compression of femtosecond pulses by second harmonic generation in the presence of substantial group velocity dispersion provides a convenient source of narrowband Raman pump pulses for femtosecond stimulated Raman spectroscopy (FSRS). We discuss here a simple and efficient modification that dramatically increases the versatility of the second harmonic spectral compression technique. Adding a spectral filter following second harmonic generation produces narrowband pulses with a superior temporal profile. This simple modification i) increases the Raman gain for a given pulse energy, ii) improves the spectral resolution, iii) suppresses coherent oscillations associated with slowly dephasing vibrations, and iv) extends the useful tunable range to at least 330-750 nm.

## 1. Introduction

Femtosecond stimulated Raman spectroscopy (FSRS) is a powerful technique for recording the structural evolution of a system following impulsive optical excitation[1]–[3]. Briefly, an ultrashort actinic laser pulse triggers a photochemical process, which is then probed via stimulated Raman scattering (SRS). SRS provides snapshots of the transient vibrational spectrum using a pair of narrowband ps Raman pump and broadband fs Raman probe pulses. Time-gating of the SRS signal based on the relative delay between the two fs pulses (actinic pump and Raman probe) reveals the evolution of the vibrational structure as a function of time following the initial excitation. Importantly, the frequency resolution of the SRS measurement is limited only by the characteristics of the narrowband Raman pump pulse (*i.e.* time and frequency profiles) and the natural line-width (*i.e.* dephasing rate) of the molecular vibration. Tuning the Raman pump pulse into resonance with an electronic transition of the target molecule provides an additional degree of molecular specificity via resonance enhancement of the SRS signal. Therefore, a key element of FSRS is the generation of spectrally narrow ps Raman pump pulses that are tunable throughout the UV-visible range.

Several methods have been used to generate suitable ps pulses from a fs laser, including direct spectral filtering of the Ti:Sapphire fundamental and second harmonic[4]. Spectral filtering alone is extremely inefficient (<0.3%) because most of the fs pulse is rejected in order to obtain ps pulses with 10-15cm$^{-1}$ bandwidth. A more efficient approach is to use non-linear frequency conversion to generate the spectrally narrow pulses. For example, Laimgruber and coworkers used sum frequency generation (SFG) with two oppositely chirped 800nm pulses to produce 400nm pulses with 12cm$^{-1}$ bandwidth and ~15% efficiency[5]. Luo and coworkers used a variation of this method based on difference frequency generation of the chirped output of a fs optical parametric amplifier (OPA) to obtain tunable ps pulses in the range 1000-1090nm (1 cm$^{-1}$ bandwidth, 0.7% efficiency)[6]. Narrowband OPAs are another efficient source of tunable ps pulses in the visible. For example, using the spectrally filtered output of a fs NOPA/OPA to seed a narrowband OPA/NOPA with chirped pumping, Shim and Mathies[7] and Co et al.[8] respectively, generated ps pulses in the range 450-780nm (10-30cm$^{-1}$ bandwidth, up to 0.1% efficiency). Kovalenko, Dobryakov, and Ernsting combined a double spectrally filtered NOPA seed with intense narrowband 400nm pump pulses from counterchirped-SFG (as described above) to significantly improve the efficiency of the narrowband NOPA in the range 450-750nm (20-30cm$^{-1}$, 3%)[9]. Similarly, Nejbauer and Radzewicz reported a narrowband OPA based on counterchirped-SFG pump and highly chirped broadband continuum seed derived from the output of a Yb laser (1030nm fundamental) to produce pulses in the range 615-985nm (10cm$^{-1}$, 4.5%)[10].

Recently, some of us demonstrated a convenient and efficient source of narrow bandwidth Raman pump pulses extending from the visible to the UV. The technique is based on spectral compression (SC) of the output of a fs OPA via second harmonic generation (SHG) in a long BBO crystal, and produces ps pulses in the range 320-520 nm with approximately 10cm$^{-1}$ bandwidth and energies reaching the multi-µJ level ~0.5% efficiency)[11]. Second harmonic-spectral compression (SH-SC) is a particularly promising technique because of the relative simplicity, efficiency, and potentially broad tunability. However, an inherent limitation of SH-SC is the unfavourable temporal profile of the narrowband pulse, as illustrated in Fig. 1. The right side of the figure shows experimental pulse shapes at three different wavelengths (measured via Kerr effect), along with exponentially-damped sine waves representing the evolution of the 802 and 1027cm$^{-1}$ vibrational modes of cyclohexane (2.0 and 0.65 ps dephasing times, respectively [12]). The sharp cut-off of the Raman pump pulse produces oscillations in each of the corresponding SRS spectra of neat cyclohexane (left side of Fig. 1) because of the Fourier relation between time and frequency domains. The spectral profile of each Raman transition is essentially the Fourier transform of the time-domain convolution of the Raman pump pulse and the evolution of the molecular vibration. The vibrational mode at 1027cm$^{-1}$ is less affected by the pulse shape because of the faster dephasing time.

In this paper, we introduce an efficient method to rectify the narrowband pulse shape using a simple spectral filter following SH-SC. Removing the wings of the pulse in the frequency domain substantially stretches and smooths the intensity profile in the time domain, with minimal energy loss compared with direct spectral filtering of a fs laser pulse. The resulting pulses are remarkably well-suited for stimulated Raman scattering.

**2. Experimental**

The experimental details of the SH-SC technique were described previously by some of us[11]. The setup is driven by an amplified Ti:Sapphire laser (Coherent Legend Elite HE) producing 3.6-mJ, 35-fs pulses at 800nm (450cm$^{-1}$ bandwidth) and 1kHz repetition rate. A two-stage OPA followed by an optional frequency conversion stage (Light Conversion TOPAS) generates tunable IR-visible pulses, which are then frequency doubled in a 25-mm BBO crystal in order to produce narrow bandwidth pulses in the UV-visible range. SH-SC takes advantage of the large group-delay-mismatch (GDM) between the fundamental and second harmonic (SH) frequencies in the long non-linear crystal.[13]–[17] Temporal walk-off as the two pulses propagate through the crystal produces a relatively long (few ps) SH pulse with compressed bandwidth relative to that of the fundamental. The SH-SC process is very efficient because SH photons are generated not only by direct SHG, but also through intra-pulse SFG among the spectral components of the fundamental that are symmetric with respect to the SHG phase-matching frequency[14], [18].

Spectral compression is a consequence of the decreasing phase-matching bandwidth as a function of crystal length. The bandwidth of the nascent SH pulse is inversely proportional to GDM in the non-linear crystal (FWHM ≈ 0.886/GDM)[19], where GDM is the product of the crystal length, $L$, and the group velocity mismatch (GVM) of the two frequencies. Therefore, the degree of spectral compression depends on both the length and identity of the non-linear crystal. For β-barium borate (BBO), the GVM of the fundamental and SH frequencies varies from 320 to 90fs/mm when tuning the fundamental from 660 to 1500nm, which gives theoretical bandwidths from 4 to 13cm$^{-1}$ for a 25mm long crystal. Indeed, SH-SC is well suited to generate multi-μJ Raman pump pulses with bandwidths <15 cm$^{-1}$ in the range 330-520nm[11].

Figure 1 shows three representative pulses obtained from SH-SC. The duration of the Raman pulse decreases with increasing wavelength due to reduced GVM at longer wavelengths. All three pulses have a strongly asymmetric "nose" shape that reflects the SHG process. The sharp trailing edge corresponds to the entrance of the fundamental into the BBO, where the SHG process is most efficient, and the roughly quadratic temporal profile reflects the decreasing SHG efficiency due to attenuation and broadening of the fundamental as the pulses propagate through the non-linear crystal. The pulse profile directly affects the vibrational lineshapes in the corresponding SRS spectra shown on the left side of the figure. Importantly, the ringing is due to the sharp cut-off in the time domain, and becomes progressively worse with increasing wavelength because of the shorter pulse duration. We emphasize that the spectral ringing is most problematic for relatively long-lived coherences, such as the 802cm$^{-1}$ mode of cyclohexane (dephasing time of 2.0 ps), and that the effect is most severe at long wavelengths, where lower GVD in the BBO crystal gives shorter pulses.

We introduce here a simple method to improve the temporal profile of pulses generated by SH-SC that also extends the useful range of the technique to significantly longer wavelengths. This is achieved by spectrally filtering the pulses obtained from SH-SC. We use a double-pass (2$f$) spectral filter with a single grating (1800lines/inch, 410nm blaze) and an adjustable slit in the collimated region of the spectrally dispersed beam[7]. The first two panels of Fig. 2 show the characteristics of a spectrally filtered 520nm pulse as a function of the slit width. Narrowing the slit suppresses the spectral wings (left panel), while the corresponding temporal profile evolves from an asymmetric "nose" shape characteristic of the unfiltered spectral compressor[14] to a broader, more symmetric profile (center panel). The right panel of Fig. 2 shows the SRS spectra of neat cyclohexane obtained with the spectrally filtered SH-SC Raman pump pulses and broadband Raman probe (Stokes) pulses derived from white-light continuum generation in a CaF$_2$ crystal[11]. Most notably, the spectral ringing of the 802cm$^{-1}$ mode decreases with slit width due to the improved temporal profile of the narrowband pulse. The spectral resolution also increases due to the longer pulse duration.

Stimulated Raman scattering is a $\chi^{(3)}$ process, where successive interactions with the Raman pump and broadband Raman probe fields generate a vibrational coherence, followed by a third interaction, also with the Raman pump, that creates the third-order polarization responsible for the SRS signal. Ideally, the Raman pump pulse intensity does not vary over the lifetime of the vibrational coherence, in which case the vibrational lineshape is simply given by the dephasing rate. However, a strongly varying intensity profile artificially alters the timescale of the third interaction, and thus influences the spectral lineshape. In the case of the sharp cut-off for the unfiltered SH-SC pulse, the Raman process is sharply gated on the last interaction and strong frequency ringing is observed for the slowly dephasing 802cm$^{-1}$ vibration of cyclohexane. The modulation is much weaker for the other vibrations in the spectrum, because those vibrations dephase more rapidly and are therefore less sensitive to the pulse structure.

The spectra in Fig. 3 further illustrate the temporal dependence of the SRS signal and the superior performance of the spectrally filtered SH-SC pulses. From left to right, the SRS spectra are obtained for progressively later arrival time of the fs Raman probe pulse. The spectra are easily explained by considering the sequence of field-matter interactions in the SRS process. For example, the delay between the first two interactions is necessarily very small (limited by the dephasing time of the electronic coherence), therefore the amplitude of the vibrational coherence induced by the second interaction is roughly proportional to the intensity of the Raman pump pulse at the time of arrival of the fs probe pulse. On the other hand, the spectral lineshape for a given transition depends on the time-evolving likelihood of the third interaction, which is essentially a convolution of the vibrational coherence with the pulse profile. The time available for the third interaction is largest when the probe arrives early, therefore the earliest delay time (left panel) gives the best frequency resolution, at the expense of modest signal strength. Conversely, the delayed arrival of the Stokes field in the other panels increases the amplitude of the vibrational coherence, but decreases the time window left to complete the $\chi^{(3)}$ process and thus reduces the resolution. In the case of the unfiltered SH-SC pulse, the third interaction is heavily gated by the sharp drop of intensity, resulting in a strong oscillatory profile in the frequency domain. Remarkably, the longer and more symmetric shape of the spectrally filtered SH-SC pulse simultaneously improves the spectral resolution, eliminates spectral ringing, and maintains strong Raman gain for all three delays shown in the bottom row of Fig. 3.

Finally, we highlight the fact that the spectral filter introduced here extends the useful range of the SH-SC method by compensating for the short pulse duration obtained at longer wavelengths due to smaller GVD in the long BBO crystal. Although the reduced SC at longer wavelengths also reduces the efficiency of the spectral filter, because more bandwidth is rejected from the pulse, it is largely compensated by the increased efficiency of the fs OPA at these wavelengths. Fig. 4 shows that excellent spectra can be recorded up to at least 700nm. The primary limitation is the efficiency of the 400nm blazed grating. In principle, longer wavelengths would also be accessible using a grating blazed for a longer wavelength. Even with the current grating, the SRS spectra in Fig. 4 are at longer wavelengths than was previously possible using SH-SC.

## 3. Conclusion

In this work we have demonstrated a simple and efficient method for controlling the temporal profile of narrowband pulses generated by SH-SC. Specifically, multi-µJ, narrow-bandwidth (~10cm$^{-1}$) pulses with symmetric time profile can be easily synthetized with broad tunability across the entire visible-near UV range (320-750nm). This technique provides an ideal source of tunable ps Raman pump pulses for FSRS, enabling resonance enhanced SRS with high frequency resolution over a wide range of wavelengths. Compared with the original SH-SC method, spectral filtering provides substantial improvements in spectral resolution and tunability, without impacting the Raman gain.

The authors are indebted to Giulio Cerullo for several enlightening discussions. E.P, C.F and T.S acknowledge support from the European Research Council under the FP7/2007-2013)/ERC IDEAS grant agreement FEMTOSCOPY Contract (No.207916). CGE is grateful for hospitality and support while at Universita di Roma "Sapienza" as a Visiting Professor, and for additional support from a National Science Foundation CAREER Award (NSF-1151555).

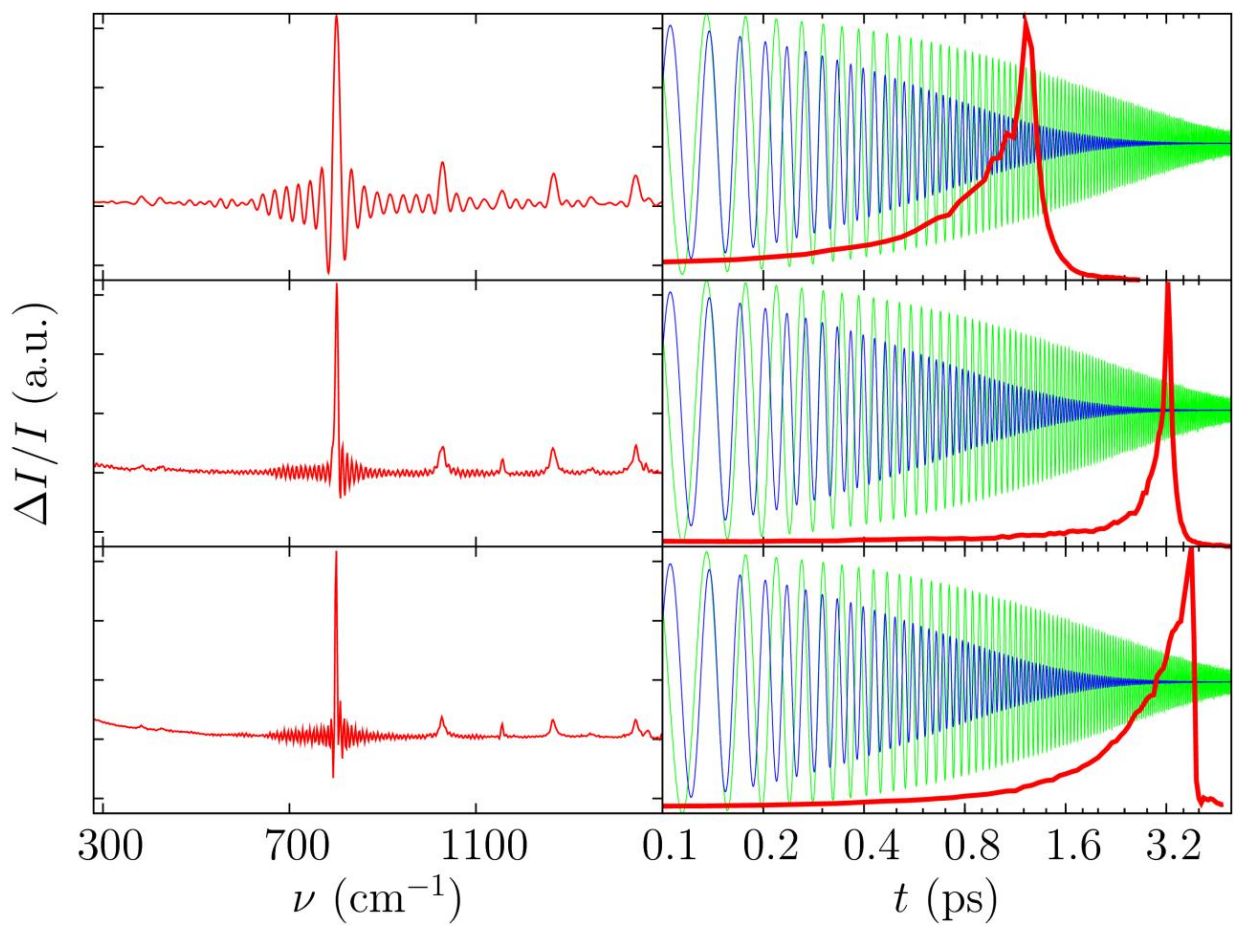

Fig. 1. Left column: Stimulated Raman spectra of cyclohexane using the standard SH-SC technique to obtain Raman pump pulses at 520nm, 430nm and 380nm (top to bottom). Right column: temporal profile of each Raman pump pulse (red) superimposed with the temporal profiles of the 802cm$^{-1}$ (green) and 1027 cm$^{-1}$ (blue) vibrational modes of cyclohexane. Time-zero is defined by arrival of the

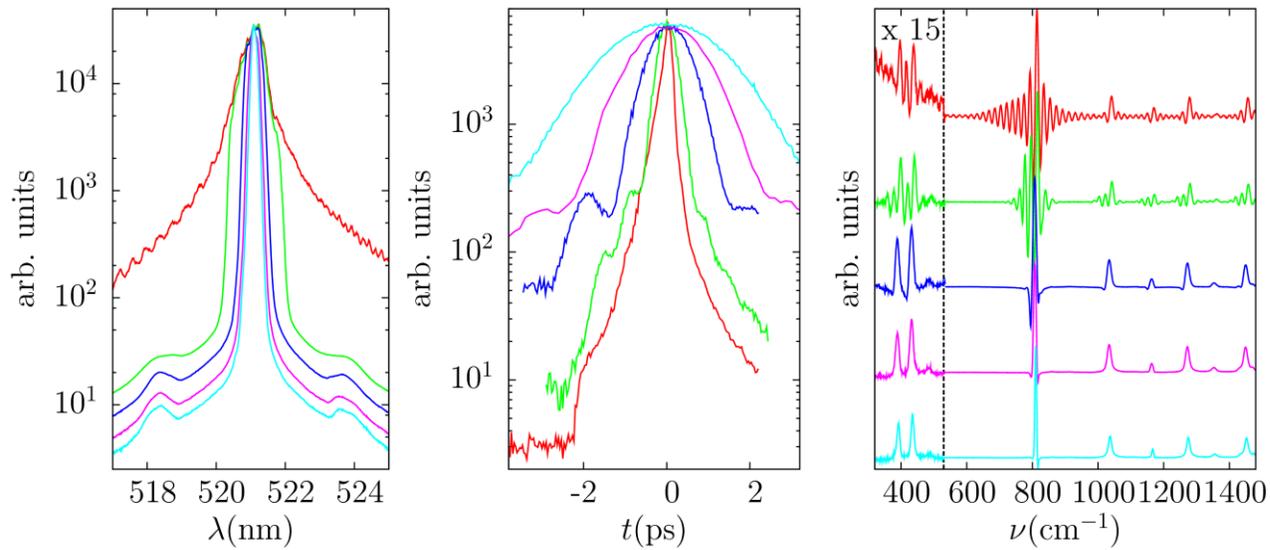

Fig. 2. Slit-width variation of a Raman pump pulse centered near 520nm. (a) Spectral profile, (b) temporal profile, and (c) corresponding stimulated Raman spectrum of cyclohexane for an open slit (red), and slit widths of 0.6 mm (green), 0.3 mm (blue), 0.2 mm (magenta), and 0.15 mm (cyan).

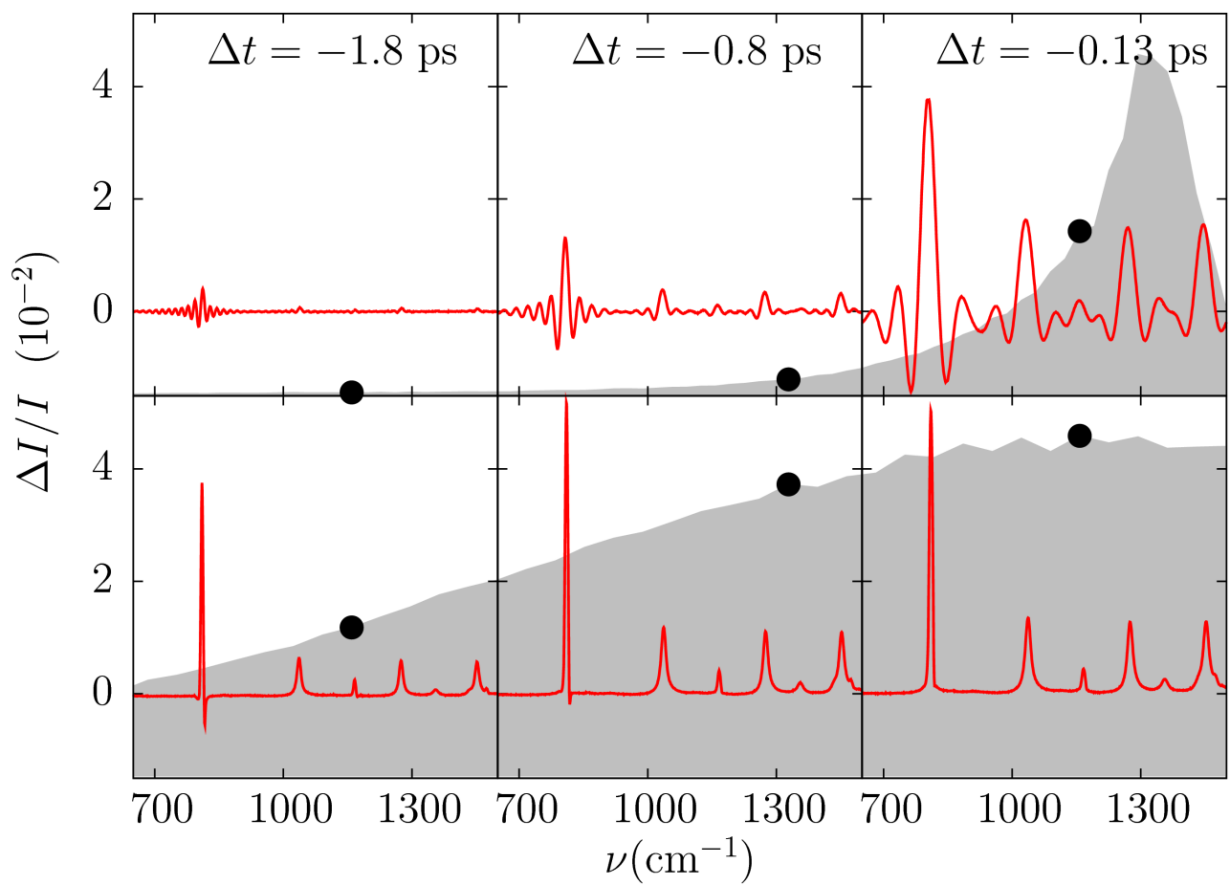

Fig. 3. Cyclohexane spectra for different relative delays between the Raman pump and probe pulses. Grey shaded areas give the temporal profile of the 520nm Raman pump for an open slit (i.e. unshaped SH-SC pulse; top panels) and for a slit width of 0.15 mm (lower panels). Black dots indicate arrival time of the femtosecond Raman probe pulse in each panel. $\Delta t$=0 is the maximum of the Raman pump pulse.

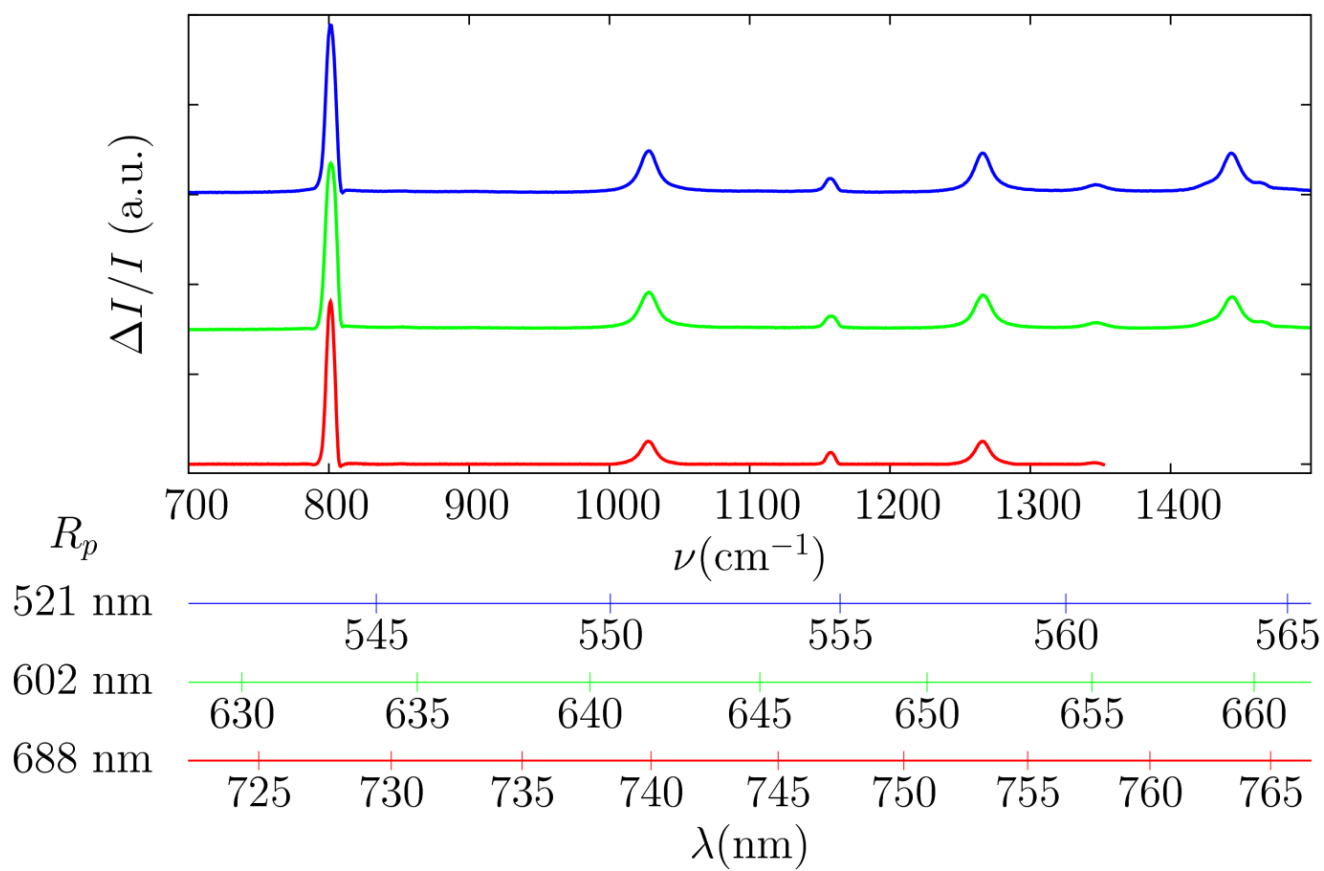

Fig. 4. Cyclohexane spectra for Raman pump wavelengths of 521, 602, and 688nm (top to bottom). Scale bars show corresponding probe wavelengths.